\documentclass{ws-ijmpc}

\begin{document}

\markboth{I. Mart\'inez-Mart\'inez \& R. L\'opez-Ruiz}
{Directed Random Markets: Connectivity determines Money}

\catchline{}{}{}{}{}

\title{DIRECTED RANDOM MARKETS:\\
CONNECTIVITY DETERMINES MONEY
}

\author{ISMAEL MART\'INEZ-MART\'INEZ}

\address{Department of Computer Science and Systems Engineering\\
Faculty of Science - University of Zaragoza\\
E-50009 Zaragoza (Spain)\\
ismael@imartinez.eu}

\author{RICARDO L\'OPEZ-RUIZ}

\address{Department of Computer Science and Systems Engineering \& BIFI\\
Faculty of Science - University of Zaragoza\\
E-50009 Zaragoza (Spain)\\
rilopez@unizar.es}

\maketitle


\begin{abstract}
Boltzmann-Gibbs distribution arises as the statistical equilibrium probability distribution of money among the agents of a closed economic system where random and undirected exchanges are allowed. When considering a model with uniform savings in the exchanges, the final distribution is close to the gamma family. In this work, we implement these exchange rules on networks and we find that these stationary probability distributions are robust and they are not affected by the topology of the underlying network. We introduce a new family of interactions: random but directed ones. In this case, it is found the topology to be determinant and the mean money per economic agent is related to the degree of the node representing the agent in the network. The relation between the mean money per economic agent and its degree is shown to be linear.

\end{abstract}

\ccode{PACS Nos.: 89.65.Gh, 89.75.Fb}

\section*{Introduction}

Agent-based modeling can be used to study systems exhibiting emergent properties which cannot be explained by aggregating the properties of the system's components.\cite{NKA} Statistical mechanics and economics share the property to analyze big ensembles where the collective behaviour is found out as a result of interactions at the microscopic level and where agent-based simulations can be applied. Many systems are studied in terms of the nature that defines their inner components while others are considered from the point of view of the interactions among the agents that can be pictured through a complex network. Plenty of information is encoded in connectivity patterns. Hierarchical structures appear in a natural way when we study societies and, according to many authors,\cite{CFL:2009} one of the milestones is to understand why and how from individuals with initial identical status, inequalities emerge. This is related to the question of hierarchy formation as a self-organization phenomenon due to social dynamics.\cite{CTDM:2002} Elitarian distributions can arise starting from a society where people initially own an equal share of economic resources, \textit{e.g.:} the exponential distribution for the low and medium income classes in western societies.\cite{DY:2001,DY:2001B,SY:2005}

We consider Dragulescu-Yakovenko gas-like models in economic exchanges\cite{Yak:2008} so, let our system be composed of $N$ economic agents, being $N\gg 1$ and constant. Each agent $i$ owns an amount of money $m_i$ so the state of the system at a given time is defined by the values that every variable $m_i$ takes at that moment, $\{m_i\}_{i=1}^N.$ Money distribution among the agents should never be confused with the notion of wealth distribution. Money is only one part inside the whole concept of wealth. Transfer of money represents payment for goods and services in a market economy. We study simplified models which keep track of that money flux but do not keep track of what goods or services are delivered. At each interacting step, agents trade by pairs and local conservation of money is sustained,
\begin{align}
(m_i,\,m_j)\longmapsto(m_i',\,m_j')\;:\;\;&m_i'=m_i+\Delta m\label{transactionDelta},\\
 &m_j'=m_j-\Delta m.\nonumber
\end{align}
Transactions result in some part of the money involved in the interaction changing its owner. For simplicity, we do not consider models where debts are allowed. 

It is deeply established in the common knowledge that highly-ranked individuals in societies have easier access to resources and better chances to compete. This is a motivation to look for internal correlations between money and surrounding environment. We wonder if the exchange rules that define simple gas-like models for random markets, when implemented on networks, are capable of depicting correlations between purchasing power of an agent inside a social network and the influence of the agent on the rest of the system. We associate the purchasing power concept to the mean money per economic agent computed as a function of the connectivity degree of each agent in a network. At this level, influence of an agent is only related to the degree of the node representing the agent. We implement the exchange rules on two type of networks: uniform random spatial graph and Barab\'asi-Albert model, and then examine the relationship between the former econo-social agent indicators for the different underlying architectures.

In section \ref{sec:randommarkets}, we review two well-known random undirected exchange rules: general and uniform savings models. In section \ref{sec:RDE}, we introduce a new family of interactions: random but directed ones. The main property of this simple exchange rule is that it is a real inspired model where social inequalities in money distribution emerge in a natural way. In section \ref{sec:Ns}, we show the relation between mean money per economic agent and the connectivity degree of the agent. For the models with undirected exchange rules, we observe no correlation between money and the degree of the nodes. Linear dependence is found for the new random exchange model we propose. Section \ref{sec:conc} is devoted to gather the most relevant conclusions.

\section{Undirected random markets}\label{sec:randommarkets}

\subsection{Undirected random market}

For some random economic systems where money is a conserved quantity, the asymptotic distribution of money among the agents is given by the Boltzmann-Gibbs distribution (BG),
\begin{equation}
p_{\text{eq}}(m_i=m)=\frac{1}{\langle m\rangle}e^{-m/\langle m\rangle},
\end{equation}
where the role of the effective temperature is played by the average amount of money per agent, 
\begin{equation}
\langle m\rangle=\frac{1}{N}\sum_{i=1}^Nm_i.
\end{equation}
This feature was first shown by Dragulescu and Yakovenko in 2000 by means of numerical simulations.\cite{DY:2000} Subsequently, analytical justification was given by L\'opez-Ruiz \textit{et al.} in 2008 and 2012. BG can be geometrically deduced\cite{LSC:2008} under the assumption of equiprobability of the possible economic microstates. We also know that an asymptotic evolution towards BG is obtained regardless of the initial distribution for those systems with total money fixed and when considering random symmetric interactions between pairs of components.\cite{LLC:2012} This comes from BG being the stable fixed point of the distributions' space, $L_1^+[0,\infty)=\{p(x)\,:\,\int_0^\infty p(x)\,dx\leq\infty\},$ under the iterated action $p(x)\rightarrow p'(x)$ of the integral operator $\mathcal{T}$ given by
 \begin{equation}
p'(z)=\big[\mathcal{T}p\big](z)=\iint_{S(z)}\frac{p(x)\,p(y)}{x+y}\;dx\,dy,\label{integral}
\end{equation}
where $S(z)=\{(x,y),\;x,y>0,\;x+y>z\}$ is the integration domain.\\

Let us now consider the gas-like model originally proposed by Dragulescu and Yakovenko so, at each computational step, we randomly choose a pair of agents and then, one -the labeled as $i$- is chosen to be the winner in the interaction process and the other one -labeled as $j$- becomes the loser and, according to the previously stated rule (\ref{transactionDelta}), an amount of money $\Delta m$ is transferred from the loser to the winner. Assuming $\Delta m\geq 0$, it is obvious that if the loser does not have enough money to pay, which is nothing but the local condition $m_j<\Delta m$, the transaction is forbidden and we should proceed with a different pair of agents.

\begin{figure}[t]
\centerline{\psfig{file=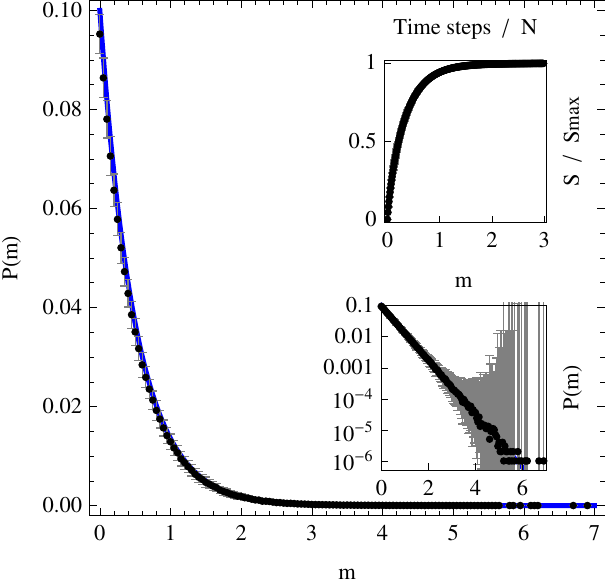,width=5.8cm}}
\vspace*{8pt}
\caption{Undirected random market. Numerical simulation considering $N=5000$ agents and \text{averaging} over $250$ samples. Every agent starts with $m_i=\langle m\rangle=0.5$ and the exchange rule is given by equation (\ref{transactionRandom}). Stationary probability distribution of money $P(m)$, computed after $N^2$ interactions, compared to the blue solid curve describing the Boltzmann-Gibbs law $P(m)\propto e^{-m/\langle m\rangle}.$ Log-linear plot and entropy time-evolution are the insets. \label{RG}}
\end{figure}

Instead of considering the restriction in the interaction, we state the exchange rule considering $\Delta m=\varepsilon(m_i+m_j)-m_i$ that gives rise to the completely random case given by 
\begin{eqnarray}
(m_i,\,m_j)&\longmapsto &(m_i',\,m_j')\;:\label{transactionRandom}\\& &\;\; m_i'=\varepsilon(m_i+m_j),\nonumber\\
& &\;\; m_j'=(1-\varepsilon)(m_i+m_j),\nonumber
\end{eqnarray}
where $\varepsilon\in[0,1]$ is a uniform random number which is refreshed at every computational step. Observe that both agents can be winner or loser in a symmetric way, depending on the random number $\varepsilon$ at each step. This approach also ensures that no agent will evolve to own a negative amount of money or, in other words, debts are not allowed. The condition $m_i\geq 0$ for every agent $i$ in the system is accomplished in a natural way. This exchange rule (\ref{transactionRandom}) is a very rough macroeconomic model where individuals or corporations raise their money for a venture and then, the market effect or their mutual interaction determines the final distribution. See figure \ref{RG}.

\subsection{Undirected random market with uniform savings}

The concept of savings arises in an obvious way from observing human behaviour when people are inmerse in a market economy.\cite{Frank} This feature is introduced through a parameter, $\lambda\in[0,1]$, which is called a propensity factor.\cite{CC:2000} This means that each agent saves a fraction $\lambda$ of its money when an interaction occurs and trades randomly with the other part:
\begin{eqnarray}
(m_i,\,m_j)&\longmapsto &(m_i',\,m_j')\;:\label{transactionSavings}\\& &\;\; m_i'=\lambda m_i+\varepsilon(1-\lambda)(m_i+m_j),\nonumber\\
& &\;\;m_j'=\lambda m_j+(1-\varepsilon)(1-\lambda)(m_i+m_j).\nonumber
\end{eqnarray}
\begin{figure}[t]
\centerline{\psfig{file=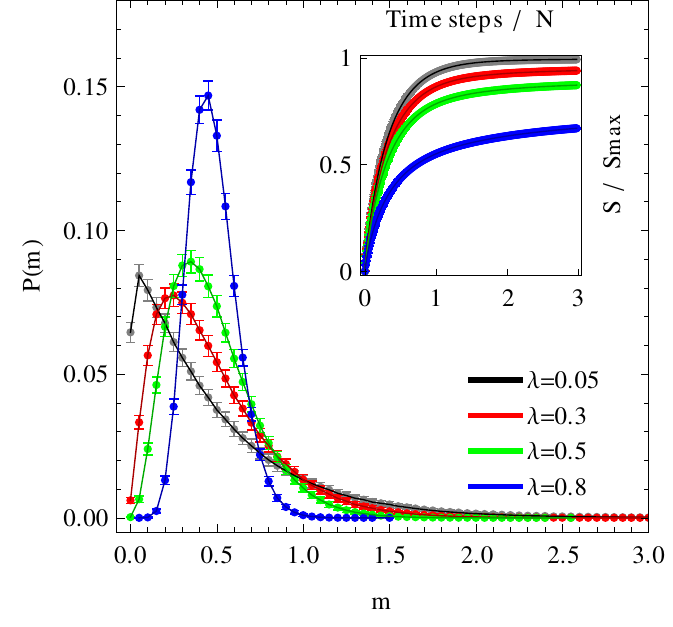,width=6.2cm}}
\vspace*{8pt}
\caption{Undirected random market with uniform savings. Numerical simulation considering $N=5000$ agents and averaging over $250$ samples. Every agent starts with $m_i=\langle m\rangle=0.5$ and the exchange rule is given by equation (\ref{transactionSavings}). Stationary probability distribution of money $P(m)$, computed after $N^2$ interactions for different values of $\lambda.$ Entropy time-evolution insets. \label{SG}}
\end{figure}
We consider the model with uniform savings which means that $\lambda$ is fixed to be constant among the agents and with no dependence on the time. The statistically stationary distribution $P(m)$ decays rapidly on both sides of the most probable value for the money per agent which, in this case, is shifted from the poorest part of the system to $\langle m\rangle$ when $\lambda\rightarrow 1.$ See figure \ref{SG}. This behaviour was already described as a self-organising feature of the market induced by self-interest of saving by each agent without any global perspective in an analogous way to the self-organisation in markets with \-restricted comodities.\cite{CPC:2001} 

First attempt to give a quantitative description for the steady distribution towards this model evolves is due to Patriarca \textit{et al.} in 2004. They stated that numerical simulations of (\ref{transactionSavings}) could be fitted to a standard Gamma distribution.\cite{PCK:2004} Subsequently (2007), Chatterjee and Chakrabarti offered a brief study of the consequences that this modelization implies and stated that as $\lambda$ increases, effectively the agents retain more of its money in any trading, which can be taken as implying that with increasing $\lambda$, temperature of the scattering process changes.\cite{CC:2007} According to their study, fourth and higher order moments of the distributions are in discrepancy with those of the Gamma family so, the actual form of the distribution for this model still remains to be found out. Calbet \textit{et al.} (2011) gave an iterative recipe to derive an analytical expression solving an integral equation.\cite{CLLR:2011} A similar expression was derived in a different way by Lallouache \textit{et al.} (2010).\cite{LAL}

\section{New scenario: directed random market}\label{sec:RDE}

We have shown how the propensity factor $\lambda$ is introduced (\ref{transactionSavings}) as a variation for the general random undirected exchange rule (\ref{transactionRandom}) showing how, from individual responsible decissions -such as saving a fraction of your money when entering an exchange market-, self-organized distributions where the mean and the mode are close arise so, richness would be quite balanced distributed among the group. A completely different scheme\cite{C:2002} that modifies the general rule (\ref{transactionDelta}) proposes a random sharing of an amount $2m_j$ (instead of $m_i+m_j$) only when $m_i>m_j$, trading at the level of the lowest economic class in the trade. This model leads to an extreme situation in which all the money in the market drifts to one agent and the rest become truely pauper. From this idea, we give a new and more general exchange rule reflecting this directed or biased orientation for the interaction and including this particular result. We propose an integral operator which is the analytical approach to this new rule in the mean-field or gas-like case and, in section \ref{sec:Ns}, we implement this rule on networks to study how it is affected when we mix directed interactions with undirected networks.

The directed exchange can be understood as a first approach to microeconomic activities where money is transferred only in one direction, similar to payments for goods. We consider the most general family of interations,
\begin{eqnarray}
(m_i,\,m_j)&\longmapsto &(m_i',\,m_j')\;:\label{transactionDirected}\\& &\;\;m_i'=\varepsilon m_i,\nonumber\\
& &\;\;m_j'=m_j+(1-\varepsilon)m_i.\nonumber
\end{eqnarray}
where $\varepsilon\in[0,1]$ is a random number chosen with uniform probability. At each time, the system is described by the probability distribution function of money when we choose one of the agents randomly, $P(m).$ In the continuous limit, we can picture the system to be the combination of two identical copies, $P_1(u)$ and $P_2(v),$ of the original system itself,
\begin{equation}
P(m)=\frac{1}{2}P_1(m)+\frac{1}{2}P_2(m).
\end{equation}
For each interaction we can consider that the two different agents $i$ and $j$, with money values being $m_i$ and $m_j$, are two realizations of picking up randomly one agent from each copy, $P_1(u=m_i)$ and $P_2(v=m_j)$ respectively. In every interaction step $(u,v)\rightarrow(u',v'),$ there is a transaction of the agents conforming the distribution $P(m)$ to a new configuration given by
\begin{equation}
P'(m)=\frac{1}{2}P'_1(m)+\frac{1}{2}P'_2(m),\label{op1}
\end{equation}
where the agents $u'$ will conform $P_1'(m)$ and $v'$ will conform $P_2'(m).$ Let us now consider the probability of a randomly chosen agent among the first copy $P_1'(u')$ owning an amount of money $u'=x$ after the interaction happened. From (\ref{transactionDirected}), it is clear that $u>x$, and as the result $u'$ is uniformly distributed in $[0,u]$ so, the probability of obtaining a certain value $x$ is given by $1/u$. The interaction of pairs $(u,v)$ in the first configuration of the system gives rise to the evolution of $P_1(u)$ to the following probability $P_1'$ of obtaining $u'=x$:
\begin{equation}
P'_1(x)=\iint_{u>x}\frac{p(u)\, p(v)}{u}\,du\,dv\label{direct1}=\int_0^\infty p(v)\,dv\;\times\,\int_{u>x}\frac{p(u)}{u}\,du=\int_{u>x}\frac{p(u)}{u}\,du.
\end{equation}

For the second copy, $P_2'(v'),$ we should consider that money of the second agent after the interaction, $v'=x,$ should be a value between the initial money it has, $v$, and the maximum possible amount of money it can have after the interaction, which will be associated to get all the money from the first agent so, $v+u.$ Again, the length of the segment $[v,u+v]$ is $u$ so, the probability of $v'$ is uniformly distributed in that segment and so, the probability to have $v'=x$ will be $1/u$. The expression for the probability $P_2'$ of having $v'=x$ in the second copy of the system results in
\begin{equation}
P'_2(x)=\iint_{v<x<u+v}\frac{p(u)\,p(v)}{u}\,du\,dv.\label{direct2}
\end{equation}

We explicitly compute the norm, 
\begin{eqnarray}
\int_0^\infty P'_1(x)\,dx&=&\int_0^\infty dx\int_{u>x}\frac{p(u)}{u}\,du=\int_0^\infty du\int_0^u \frac{p(u)}{u}\,dx=\int_0^\infty p(u)\,du\label{normop}=1,\hspace{8mm}\\
\int_0^\infty P'_2(x)\,dx&=&\int_0^\infty dx\int_{v<x<u+v}\frac{p(u)\,p(v)}{u}\,du\,dv\nonumber\\&=&\int_0^\infty du\int_0^\infty dv\int_v^{u+v}\frac{p(u)\,p(v)}{u}\,dx=\int_0^\infty p(u)\,du\int_0^\infty p(v)\,dv=1,\nonumber
\end{eqnarray}
and the expected value,
\begin{eqnarray}
\int_0^\infty x\,P_1'(x)\,dx&=&\int_0^\infty x\,dx\iint_{u>x}\frac{p(u)}{u}\,du=\int_0^\infty du\int_0^u x\,\frac{p(u)}{u}\,dx\label{expop}\\&=&\frac{1}{2}\int u\,p(u)\,du=\frac{1}{2}\langle u \rangle.\nonumber\\
\int_0^\infty x\,P_2'(x)\,dx&=&\int_0^\infty x\,dx\iint_{v<x<u+v}\frac{p(u)\,p(v)}{u}\,du\,dv\nonumber\\
&=&\int_0^\infty du\int_0^\infty dv \int_v^{u+v}x\,\frac{p(u)\,p(v)}{u}\,du\,dv\nonumber\\&=&\int_0^\infty du\int_0^\infty \frac{1}{2}\frac{u^2+2uv}{u}\,p(u)\,p(v)\,dv\nonumber\\
&=&\frac{1}{2}\Bigg[\int_0^\infty du\int_0^\infty u\,p(u)\,p(v)\,dv+\int_0^\infty du\int_0^\infty 2\,v\,p(u)\,p(v)\,dv \Bigg]=\frac{3}{2}\langle u\rangle.\nonumber
\end{eqnarray}

From results (\ref{normop}) and (\ref{expop}), together with definition (\ref{op1}), the norm and mean value of the distributions are conserved when we define the operator for the directed random interaction by
\begin{equation}
\big[\mathcal{T}p\big](x)=\frac{1}{2}\int_{u>x}\frac{p(u)}{u}\,du+\frac{1}{2}\iint_{v<x<u+v}\frac{p(u)\,p(v)}{u}\,du\,dv.\label{opdef}
\end{equation}

\begin{figure}[t]
\centerline{\psfig{file=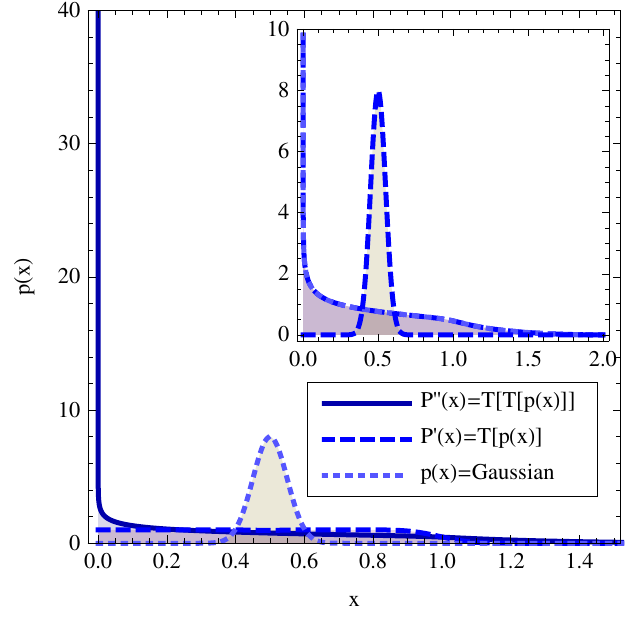,width=6.cm}\hspace{0.5cm}\psfig{file=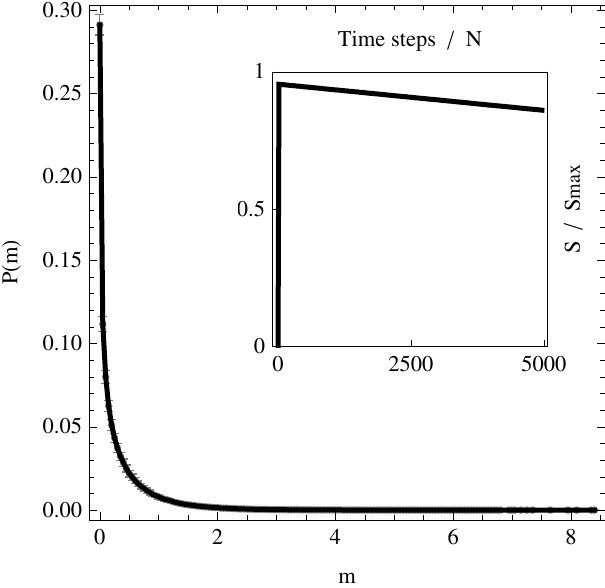,width=6.cm}}
\vspace*{8pt}
\caption{LEFT: First and second application of the integral operator (\ref{opdef}) over a Gaussian distribution. Dotted curve is the initial Gaussian distribution $N(\langle m\rangle=0.5,\,\sigma=0.05).$ The distribution resulting from the first application of (\ref{opdef}), $\big[\mathcal{T}p\big](x)$, is plotted in dashed lines. Second iteration, $\big[\mathcal{T}\big[\mathcal{T}p\big]\big](x)$, is plotted in solid line. Due to the scale proportions, inset figure shows a detailed comparison between initial distribution and the resulting from the first iteration. RIGHT: Directed random market implemented as gas-like model. Numerical simulation considering $N=5000$ agents and averaging over $250$ samples. Every agent starts with $m_i=\langle m\rangle=0.5$ and the exchange rule is given by equation (\ref{transactionDirected}). Stationary probability distribution of money $P(m)$, computed after $N^2$ interactions and inset showing the whole entropy-evolution. We can appreciate how the system starts evolving towards the state of maximum entropy and then it starts a self-organising process which is very slow. When considering extremely long runs of the simulation we observe how the entropy falls  towards states of negligible entropy, as we should expect from a system where lack of information in $P(m)$ is just the permutation of the agent owning all the money. \label{operador}}
\end{figure}

Although we cannot give a proof for the infinite iteration of this operator, we see that it piles up the distribution at the lower values for $m$ which gives rise to a very impoverished population and a very slightest fraction of too opulent agents. Even with such a very narrow initial distribution we choose, the effect of this operator is very strong in only a couple of iterations. This result is in perfect agreement with the first model of biased interaction\cite{C:2002} we mentioned in our previous section. Other models with asymetric rules that establish a transition between Boltzmann-Gibbs and Pareto distributions can be found in the literature.\cite{PLLR11} See figure \ref{operador}.

\section{Simulations on networks}\label{sec:Ns}

We stated in the introduction of this work the desire of finding a first model which, when implemented on networks, is able to show a relation between money and influence of an agent on the system. For this purpose, we simulate the two models we have already studied in section \ref{sec:randommarkets} and the new one we have presented in section \ref{sec:RDE}. We choose two representative cases: the random uniform spatial network (SP) and the Barab\'asi-Albert model (BA). SP is an easy way to build networks with Poissonian degree distributions\cite{SP} and BA is an algorithm for generating random scale-free networks following the preferential attachment prescription.\cite{BA} 

At the beginning, every agent is given the same amount of money so, the initial distribution of money among the agents is written
\begin{equation}
P_{\text{initial}}(m)=\delta(m-\langle m\rangle),
\end{equation}
and we obtain the steady state distribution for the exchange rules (\ref{transactionRandom}), (\ref{transactionSavings}) and (\ref{transactionDirected}) implemented on these two different topologies. From the histogram related to the distribution of money among the agents, we can consider the entropy associated to that distribution in a discrete form,
\begin{equation}
S=-\sum_{i=1}^{I_{max}}P_\Delta (i)\,\log P_\Delta(i),
\end{equation}
where $\Delta$ just recalls the coarse graining when computig $P(m)$ as a discrete histogram. For simplicity, we measure $S/S_{\text{max}},$ the entropy as a fraction of the maximum which is given for the exponential distribution.

\begin{figure}[t]
\vspace{0.2cm}
\begin{center}(a) Spatial network:\end{center}
\centerline{\psfig{file=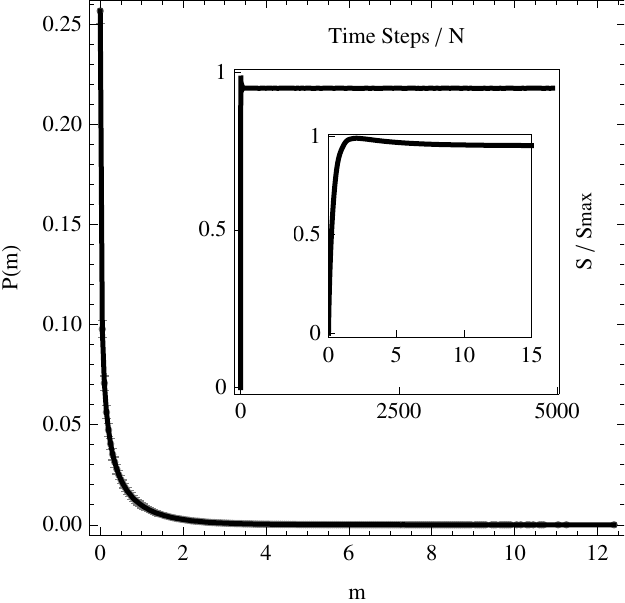,width=6.cm}\hspace{0.5cm}\psfig{file=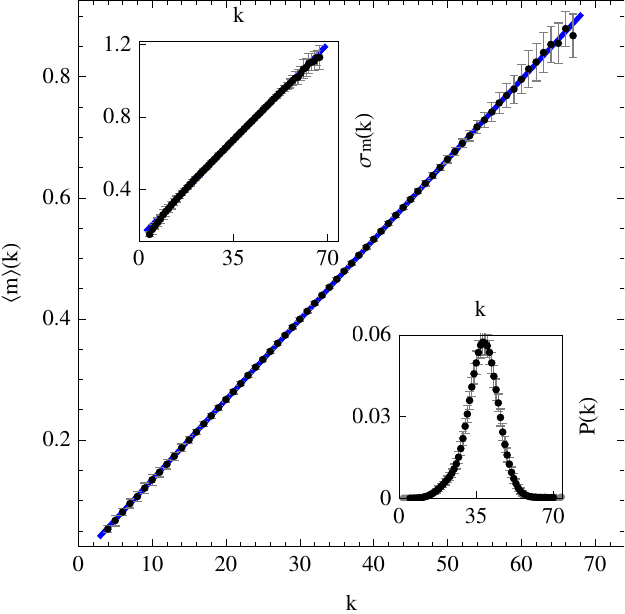,width=6.cm}}
\begin{center}(b) Barab\'asi-Albert model:\end{center}
\centerline{\psfig{file=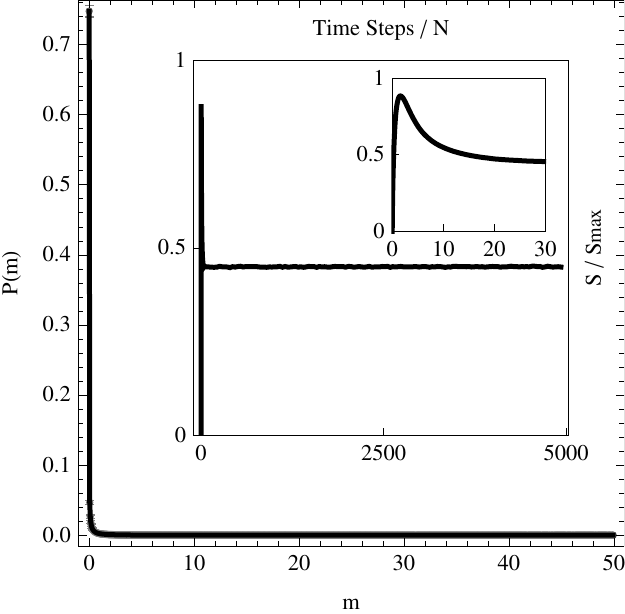,width=6.cm}\hspace{0.5cm}\psfig{file=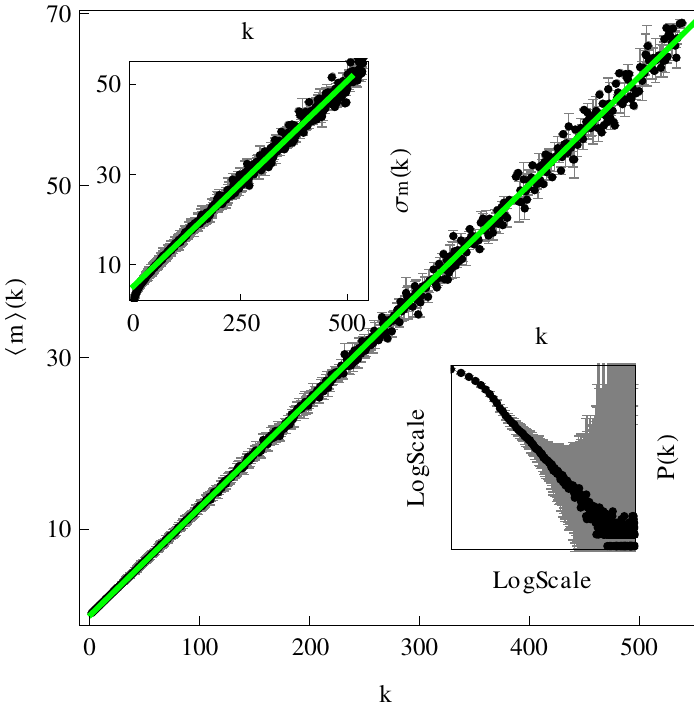,width=5.9cm}}
\vspace*{8pt}
\caption{Directed random market implemented on networks. Numerical simulation considering $N=5000$ agents and averaging over $250$ networks for each type (a) and (b). Every agent starts with $m_i=\langle m\rangle=0.5$ and the exchange rule is given by equation (\ref{transactionDirected}). LEFT: Stationary probability distribution of money $P(m)$, computed after $N^2$ interactions and entropy-evolution for the whole time of the simulation with inset showing transitory regime at the beginning. We can appreciate how the system evolves towards a highly entropic state and then it start a self-organising process which is analogous to what we found for the gas-like approach in figure \ref{operador} but, now, it reaches a stable regime. The steady state is characteristic for each network. This comes from the different values at which $S/S_{max}$ stabilizes. RIGHT: Mean money per economic agent computed according to (\ref{mmk}) and its standard deviation, given by (\ref{sigmamk}). Linear fit is obvious. Typical degree distribution is also shown. \label{DNet}}
\vspace{0.2cm}
\end{figure}

If the money distribution is affected by the topology of the underlying network, it should show some kind of dependence on the degree of the agents. For our purposes it is enough for the reader to associate the degree of a node to its number of neighbouring agents it can interact with. As we will only use for now undirected and static networks, the degree of an agent will be a distinctive feature. We define the \textit{mean money per economic agent as a function of k}, $\langle m\rangle(k),$ given by
\begin{equation}
\langle m\rangle(k)=\frac{1}{N^2N_k}\sum_{\tau=1}^{N^2}\sum_{i=1}^{N_k}m_i(\tau)\;\;\forall\,i\;:\;k_i=k.\label{mmk}
\end{equation}
Note that we compute the mean money of the nodes inside each class of connectivity at every step of our simulation and then, we consider the time-averaged mean money per node according to the different degrees. $N_k$ denotes how many nodes have degree $k$ and our simulations run for $N^2$ steps. There is no need to be worried about the transitory regime disturbing the computed results for $\langle m\rangle (k)$ because it is negligible: less than $0.5\%$ of total computed steps in the worst case (see figure \ref{DNet}). We also compute the standard deviation, $\sigma_m(k)$, given by
\begin{eqnarray}
\sigma_m(k)=\sqrt{\frac{1}{N^2N_k}\sum_{\tau=1}^{N^2}\sum_{i=1}^{N_k}\Big[m_i(\tau)-\langle m\rangle(k)\Big]^2}\label{sigmamk}\\\nonumber\;\;\forall\,i\;:\;k_i=k.
\end{eqnarray} 

\begin{figure}[t]
\vspace{0.2cm}
\begin{center}(a) Spatial network:\end{center}
\centerline{\psfig{file=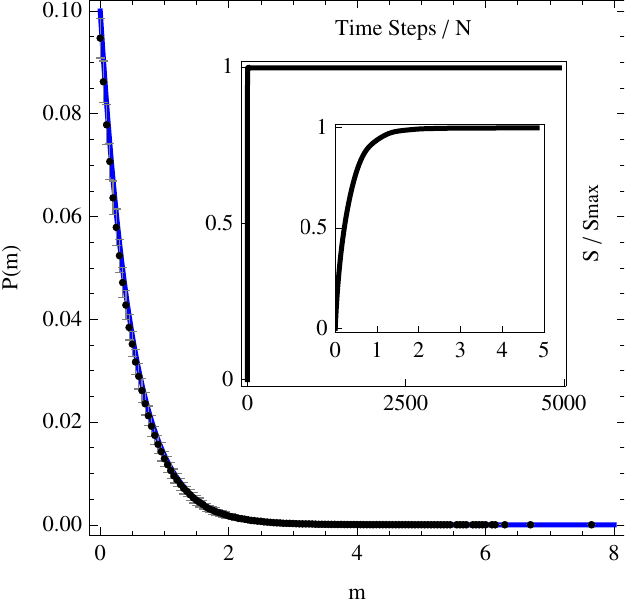,width=6.cm}\hspace{0.5cm}\psfig{file=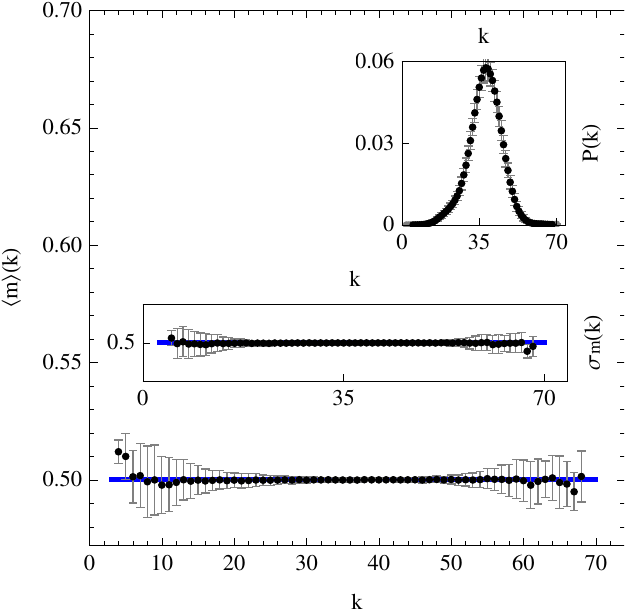,width=6.cm}}
\begin{center}(b) Barab\'asi-Albert model:\end{center}
\centerline{\psfig{file=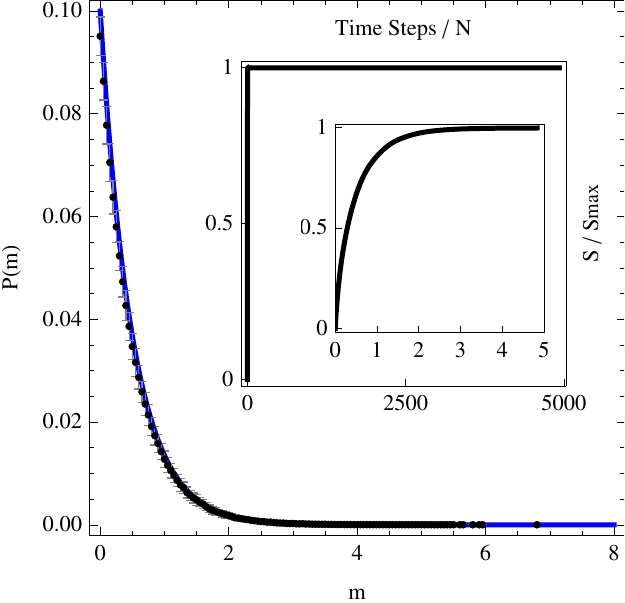,width=6.cm}\hspace{0.5cm}\psfig{file=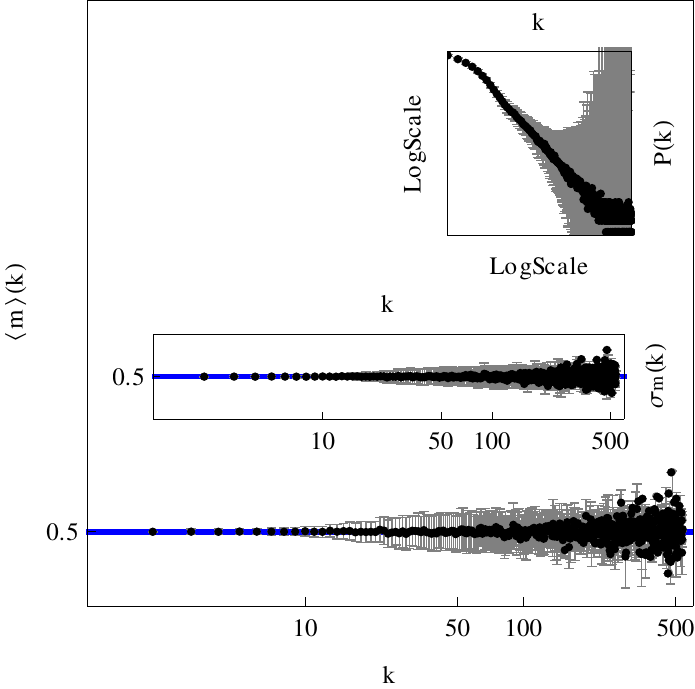,width=6.cm}}
\vspace*{8pt}
\caption{Undirected random market implemented on networks. Numerical simulation considering $N=5000$ agents and averaging over $250$ networks for each type (a) and (b). Every agent starts with $m_i=\langle m\rangle=0.5$ and the exchange rule is given by equation (\ref{transactionRandom}). LEFT: Stationary probability distribution of money $P(m)$, computed after $N^2$ interactions, compared to the blue solid curve describing the Boltzmann-Gibbs law $P(m)\propto e^{-m/\langle m\rangle}$ and entropy-evolution for the whole time of the simulation and inserted detail for the initial transitory. RIGHT: Mean money per economic agent computed according to (\ref{mmk}) and its standard deviation, given by (\ref{sigmamk}). Typical degree distribution is also shown. \label{RNet}}
\vspace{1.2cm}
\end{figure}

\begin{figure}[t]
\vspace{0.2cm}
\begin{center}(a) Spatial network:\end{center}
\centerline{\psfig{file=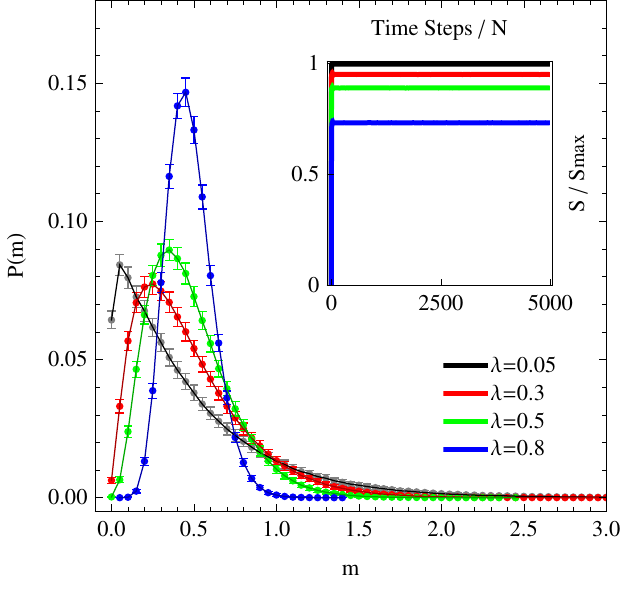,width=6.3cm}\hspace{0.5cm}\psfig{file=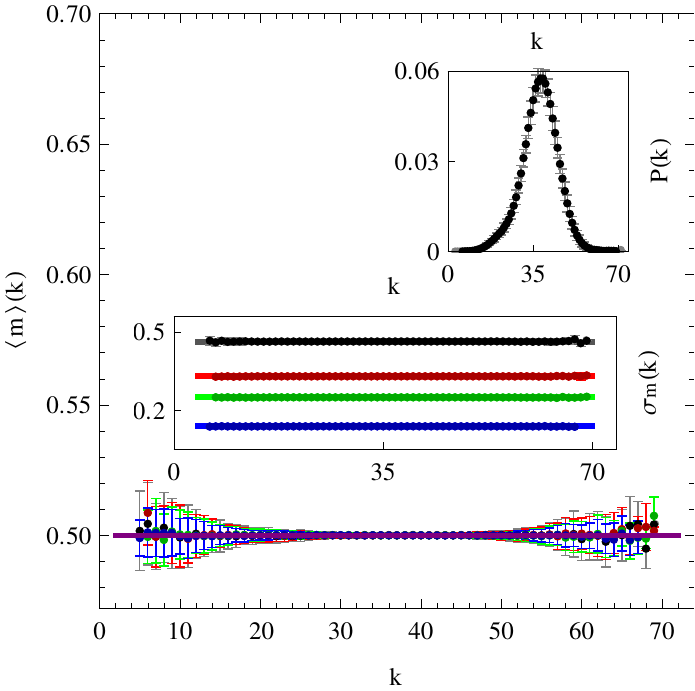,width=6.cm}}
\begin{center}(b) Barab\'asi-Albert model:\end{center}
\centerline{\psfig{file=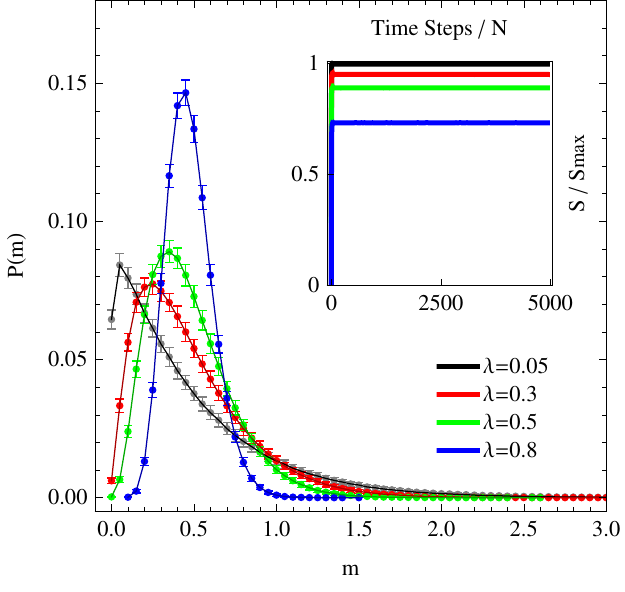,width=6.3cm}\hspace{0.5cm}\psfig{file=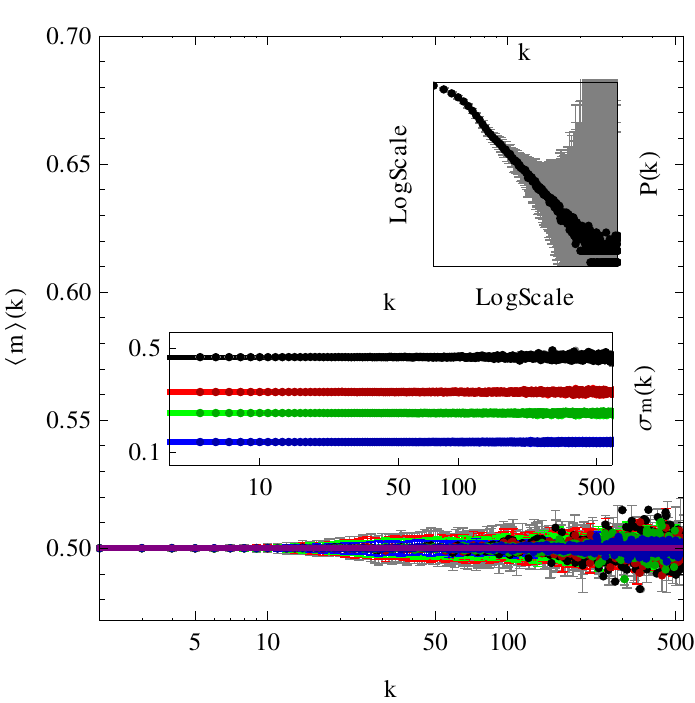,width=6.2cm}}
\vspace*{8pt}
\caption{Undirected random market with uniform savings implemented on networks. Numerical simulation considering $N=5000$ agents and averaging over $250$ networks for each type (a) and (b). Every agent starts with $m_i=\langle m\rangle=0.5$ and the exchange rule is given by equation (\ref{transactionSavings}). LEFT: Stationary probability distribution of money $P(m)$, computed after $N^2$ interactions for different values of $\lambda$ and entropy-evolution for the whole time of the simulation. Transitory towards equilibrium entropy is analogous to the one shown in figure \ref{RNet}. RIGHT: Mean money per economic agent computed according to (\ref{mmk}) and its standard deviation, given by (\ref{sigmamk}). Typical degree distribution is also shown. \label{SNet}}
\vspace{1.2cm}
\end{figure}

In figures \ref{RNet}, \ref{SNet} and \ref{DNet} we show, for the rules (\ref{transactionRandom}), (\ref{transactionSavings}) and (\ref{transactionDirected}), respectively, the stationary probability distribution $P(m)$ with entropy-evolution of $S/S_{\text{max}}$ for the whole simulation with inset detail of the transitory regime when it is required. When we plot the distributions $\langle m\rangle(k)$ and $\sigma_m(k)$, it is also shown the characteristic degree distribution of the networks we implement. 

We clearly see in these figures how rules (\ref{transactionRandom}) and (\ref{transactionSavings}) are transparent to the underlying topology, provoking always the decay of the economic system to the BG or Gamma-like distributions, respectively, indistinctly of the different connectivities of the agents. We can say that for these types of interactions the economic classes are blind respect to the social influence of the agents. By contrary, the new directed rule (\ref{transactionDirected}) separates the agents in economic classes correlated with their connectivities, in such a way that more connected agents show a bigger propensity to accumulate more money, in this case with a linear relationship between money and connectivity. This is a characteristic that, in general and independently of the political system installed in the power, seems to be more likely to be found in the reality.

\newpage

\newpage

\newpage

\newpage
\section{Conclusion}\label{sec:conc}

We have found that topology does not determine the final equilibrium distribution for the family of undirected random markets, both with or without uniform savings, as can be seen by comparing the distributions $P(m)$ from figures \ref{RNet} and \ref{SNet} to those from figures \ref{RG} and \ref{SG}. Thus, from the uniform value of $\langle m\rangle(k)$ and $\sigma_m(k)$, we can conclude that the connectivity of an agent immersed in undirected random markets does not determine if the agent will have more or less money, that is, the degree $k$ of an agent does not decide its richness. 

When we consider the directed random market given by the operator (\ref{opdef}), simulations plotted in the figure \ref{DNet} suggest that this model reproduces very clearly the real insight based on the impression that in certain societies the richness is owned by a very small fraction of the population and poverty is extended among the majority of the agents. For this type of economies, we also discover that the underlying topology determines the stationary distribution of money among the agents, $P(m)$, and that the connectivity of each agent proportionally determines its average richness.

Assuming that we can apply this rule for any quantity that can be exchanged, and not only money, this result introduces the interesting idea of how we can create systems with a property shared by its inner components with an steady distribution essentially determined by the topology defined by the connections between the agents, although evidently this statistical equilibrium is dynamical and presents a continuous flow between those agents. It is also interesting to highlight how this system evolves from the initial zero-entropy state to a state with maximum entropy and then how it relaxes towards the asymptotic equilibrium state.

Let us conclude by saying that we have introduced a new directed random market model in the context of economic gas-like models, which can be understood as a first and simple model characterized by {\it more connectivity implies more money}.

\end{document}